\documentclass[twocolumn,english,aps,superscriptaddress,groupedaddress,amsmath,amssymb,floatfix]{revtex4}
\usepackage[]{fontenc}
\usepackage[latin9]{inputenc}
\usepackage{amsmath}
\usepackage{graphicx}
\usepackage{amssymb}
\usepackage{esint}

\makeatletter
\@ifundefined{textcolor}{}
{%
 \definecolor{BLACK}{gray}{0}
 \definecolor{WHITE}{gray}{1}
 \definecolor{RED}{rgb}{1,0,0}
 \definecolor{GREEN}{rgb}{0,1,0}
 \definecolor{BLUE}{rgb}{0,0,1}
 \definecolor{CYAN}{cmyk}{1,0,0,0}
 \definecolor{MAGENTA}{cmyk}{0,1,0,0}
 \definecolor{YELLOW}{cmyk}{0,0,1,0}
 }

\usepackage{times}

\newcommand{\Tr}{\mathrm{Tr}}

\newcommand{\tr}{\mathrm{tr}}
\makeatother

\usepackage{babel}

\makeatother

\usepackage{babel}

\makeatother

\usepackage{babel}

\begin{document}

\title{Local quenches in frustrated quantum spin chains: global vs.~subsystem
equilibration}

\author{Mathias Diez}

\affiliation{Department of Physics and Astronomy and Center for Quantum Information
Science \& Technology, University of Southern California, Los Angeles,
California 90089-0484, USA}

\affiliation{Department of Physics, University of Konstanz, D-78457 Konstanz,
Germany}

\author{Nicholas Chancellor}

\affiliation{Department of Physics and Astronomy and Center for Quantum Information
Science \& Technology, University of Southern California, Los Angeles,
California 90089-0484, USA}

\author{Stephan Haas}

\email{shaas@usc.edu}

\affiliation{Department of Physics and Astronomy and Center for Quantum Information
Science \& Technology, University of Southern California, Los Angeles,
California 90089-0484, USA}

\author{Lorenzo Campos Venuti}

\affiliation{Institute for Scientific Interchange (ISI), Viale S. Severo 65, I-10133
Torino, Italy }

\author{Paolo Zanardi}

\affiliation{Department of Physics and Astronomy and Center for Quantum Information
Science \& Technology, University of Southern California, Los Angeles,
California 90089-0484, USA}

\affiliation{Institute for Scientific Interchange (ISI), Viale S. Severo 65, I-10133
Torino, Italy }
\begin{abstract}
We study the equilibration behavior following local quenches, using
frustrated quantum spin chains as an example of interacting closed
quantum systems. Specifically, we examine the statistics of the time
series of the Loschmidt echo, the trace distance of the time-evolved
local density matrix to its average state, and the local magnetization.
Depending on the quench parameters, the equilibration statistics of
these quantities show features of good or poor equilibration, indicated
by Gaussian, exponential or bistable distribution functions. These
universal functions provide valuable tools to characterize the various
time-evolution responses and give insight into the plethora of equilibration
phenomena in complex quantum systems.
\end{abstract}
\maketitle

\section{Introduction}

The equilibration properties of closed quantum systems have been a
topic of growing interest. As opposed to equilibration in open quantum
systems coupled to an infinite bath, a closed finite quantum system
does not admit asymptotic fixed points, unless the initial state is
an eigenstate of the evolution Hamiltonian. Except for this trivial
case, the evolution state $\rho\left(t\right)$ oscillates without
converging to any limiting point. On the other hand, equilibration
in closed quantum systems may be defined by resorting to probabilistic
notions \cite{von_neumann1929,tasaki98,popescu06,linden09,reimann08,goldstein2010}.
Specifically, one says that an observable equilibrates if it remains close to its
average for most of the times along the evolution trajectory. Alternatively,
equilibration takes place if the fluctuations around the average are
small. For \emph{typical }initial states, the authors of \cite{popescu06,linden09}
proved a number of bounds on the fluctuations that imply equilibration,
provided the subsystem is small compared to the environment. Here
``typical" means ``taken uniformly at random from the space of pure states".
The insightful results in \cite{popescu06,linden09} leave nonetheless
open the question of equilibration for a number of interesting cases.
Namely when sizes are such that the environment cannot be considered
overwhelmingly larger than the subsystems and/or when the initial
state cannot be considered typical. Both such features are present
in the experimentally relevant situation of small quenches on a finite
system \cite{roux2010,biroli09}. This is the situation studied in
this article. More precisely, we suddenly change the evolution Hamiltonian
only on a subsystem $S$ and then monitor equilibration features of
the subsystem as well as of the total system by studying the full
statistics of global (i.e.~with support on the whole system) as well
as of local observables. For the local subsystem we choose the subsystem
magnetization as a physically natural observable. To obtain information
on all possible observables defined on $S$ we monitor the trace distance
of the subsystem density matrix from its equilibrium value. As a global
observable we take the Loschmidt echo (LE). The Loschmidt echo has
the additional appealing features that its time averages provides
a bound on the fluctuations on any sufficiently small subsystem. However,
in the situations under study most of the bounds are too loose and
equilibration must be checked by an explicit computation.

To be more specific, the idea is the following. We prepare the system
at $t=0$ in the ground state of the Hamiltonian $H_{0}$. At $t>0$
we suddenly modify the Hamiltonian in the region $S$. The system
then evolves without further disturbance with Hamiltonian $H=H_{0}+V_{S}$,
where the term $V_{S}$ only acts on subsystem $S$. The equilibration
properties are then evaluated by monitoring local quantities of $S$
as well as global quantities of the entire system. Throughout this
paper we will use the terms {}``local\char`\"{} resp.~{}``global\char`\"{}
for quantities which refer to subsystem $S$ resp.~the entire system.
The idea is to study the interplay between equilibration of these
{}``local" and {}``global" quantities.

After a sufficiently long evolution time, the trajectory $\rho\left(t\right)$
defines an emerging equilibrium average ensemble given by $\overline{\rho}:=\overline{\rho\left(t\right)}:=\lim_{T\to\infty}T^{-1}\int_{0}^{T}\rho\left(t\right)$,
where the observation time $T$ has been sent to infinity. The average
density matrix $\overline{\rho}$, in the eigenbasis of the evolution
Hamiltonian $H$, has the form of a dephased state: $\overline{\rho}=\sum_{n}p_{n}|n\rangle\langle n|,$
where $p_{n}=\left|\langle n|\psi_{0}\rangle\right|^{2}$ ($|\psi_{0}\rangle$
initial state), and for simplicity we assume here that the eigenenergies
$E_{n}$ are non-degenerate \cite{reimann08,linden09}. In the same
fashion, if we monitor an observable $O$ over a long time, its average
value will be given by $\overline{O}:=\overline{\langle O\left(t\right)\rangle}:=\overline{\tr\left[\rho\left(t\right)O\right]}$.
In finite systems we can safely bring the time average inside the
trace and obtain $\overline{O}=\tr\left[\overline{\rho}O\right]$.
The equilibration properties of an observable are encoded in its variance,
or more precisely in its full statistics. The quantity $\langle O\left(t\right)\rangle$
is seen as a stochastic variable with the uniform measure $dt/T$
over the observation time interval $\left[0,T\right]$ ($T$ will
be sent to infinity). We can say that the observable $O$ has good
equilibration features if its probability distribution $P_{O}$ shows
concentration around its mean value. Explicitly, the probability distribution
$P_{O}$ is given by $P_{O}\left(x\right)=\overline{\delta\left(\langle O\left(t\right)\rangle-x\right)}$.
To fix ideas further, we can say that good equilibration corresponds
to a small variance $\overline{\Delta O^{2}}$, or more precisely
to a large signal to noise ratio $\overline{O}/\sqrt{\overline{\Delta O^{2}}}$.
Clearly $P_{O}$ depends on the observable $O$ but also on the initial
state $|\psi_{0}\rangle$ and on the evolution Hamiltonian $H$.

\section{Preliminaries}

The model we study here is the frustrated antiferromagnetic Heisenberg
spin-1/2 chain, with a magnetic field acting only on a subsystem $S$
containing $N_{S}$ adjacent sites. Its Hamiltonian is given by\begin{equation}
H=\sum_{j=1}^{N}\left(J_{1}\boldsymbol{S}_{j}\cdot\boldsymbol{S}_{j+1}+J_{2}\boldsymbol{S}_{j}\cdot\boldsymbol{S}_{j+2}\right)-h_{S}\sum_{j=1}^{N_{S}}S_{j}^{z},\label{eq:hamiltonian}\end{equation}
 where $J_{1}$ and $J_{2}$ denote the nearest-neighbor and next-nearest-neighbor
couplings respectively, $J_{1}$ is set to 1 hereafter, $\boldsymbol{S}_{j}$
are spin-1/2 operators, and periodic boundary conditions are applied
throughout, i.e.~$\boldsymbol{S}_{j+N}=\boldsymbol{S}_{j}$. The
initial, i.e. pre-quench, Hamiltonian $H_{0}$ is given by Eq.~(\ref{eq:hamiltonian})
with $h_{S}=h_{S}^{\left(i\right)}$. The local field $h_{S}$ is
then suddenly changed, such that the evolution Hamiltonian $H$ is
given by Eq.~(\ref{eq:hamiltonian}) with $h_{S}=h_{S}^{\left(f\right)}$.
The couplings $J_{1},\, J_{2}$ are held fixed during the quench process.

If we consider the hard-core boson mapping of the spin variables given
by $S_{j}^{z}=b_{j}^{\dagger}b_{j}-1/2$, the local field $h_{S}$
is equivalent to a local chemical potential which confines the particles
in region $S$. The Hamiltonian Eq.~(\ref{eq:hamiltonian}) commutes
with the total magnetization $S_{\mathrm{tot}}^{z}$. Since in the
numerical studies discussed below the initial field $h_{S}$ is not
chosen excessively large, we verified that the ground state of $H_{0}$
(and so the evolved state too) always belongs to the sector with zero
total magnetization $S_{\mathrm{tot}}^{z}=0$. This means that the
system has $N/2$ hard core bosons which initially prefer to occupy
the region $S$; still the hard-core constraint ($(b_{j}^{\dagger})^{2}=0$)
does not permit to have more than one particle at the same site %
\footnote{Note also that when the initial state is in the zero magnetization
sector, evolving with a local field $h_{S}$ acting on $S$ has the
same effect as evolving with field $-h_{S}$ acting on the complementary
of $S$, $\overline{S}$. In this case the evolved state is $|\psi\left(t\right)\rangle=\exp\left[-it\left(H_{0}-h_{S}S_{S}^{z}\right)\right]|\psi_{0}\rangle=\exp\left[-it\left(H_{0}-h_{S}S_{S}^{z}\right)\right]\exp\left[-ith_{S}S_{\mathrm{tot}}^{z}\right]|\psi_{0}\rangle$.
The result follows noting that $S_{S}^{z}-S_{\mathrm{tot}}^{z}=-S_{\overline{S}}^{z}$. %
}.

The local field also breaks translational invariance. This enlarges
the dimension $d_{0}$ of the ground state manifold, i.e.~the vector
space which contains the starting state, $|\psi_{0}\rangle$ and the
evolved state $|\psi\left(t\right)\rangle=\exp\left(-itH\right)|\psi_{0}\rangle$.
This feature is important. In fact, $d_{0}$ plays the role of an
effective Hilbert space dimension. In systems with many symmetries,
$d_{0}$ can be significantly reduced with respect to the total Hilbert
space dimension. As a consequence, a system with large $d_{0}$ has
comparatively smaller finite size effects compared to a system with
the same number of sites but more symmetries and smaller $d_{0}$.

To obtain information about the system's equilibration properties
we consider various time dependent observables $f\left(t\right)$
and compute their statistics using the uniform measure $dt/T$ over
the interval $\left[0,T\right]$, in other words we compute the probability
distribution $P_{f}$ of the random variable $f\left(t\right)$ via
$P_{f}\left(x\right):=\overline{\delta\left(f\left(t\right)-x\right)}$.
We are mainly interested in the concentration properties of $P_{f}$,
that is its {}``peakedness''.

To monitor what happens to subsystem $S$ we first consider its magnetization
$m_{S}\left(t\right):=\langle S_{S}^{z}\left(t\right)\rangle/N_{S}=\langle\psi\left(t\right)|S_{S}^{z}|\psi\left(t\right)\rangle/N_{S}$.
As for any other observable $O$, it can be expressed in the basis
of the eigenvectors of the evolution Hamiltonian in the following
way:\begin{equation}
\langle O\left(t\right)\rangle=\overline{O}+\sum_{n>m}2\langle n|O|m\rangle c_{m}c_{n}\cos\left[t\left(E_{m}-E_{n}\right)\right].\label{eq:observable}\end{equation}
 Here $c_{n}=\langle n|\psi_{0}\rangle$, and we assumed that both
the observable $O$ and the initial state $|\psi_{0}\rangle$ can
be chosen real in the basis of $H$.

The local magnetization $S_{S}^{z}$, although being quite natural,
is just one of the $2^{N_{S}}$ linearly independent observables which
we can define on $S$. To obtain information on all possible observables
on $S$ one has to consider the density matrix of the subsystem $S$:
$\rho_{S}\left(t\right)=\tr_{\overline{S}}|\psi\left(t\right)\rangle\langle\psi\left(t\right)|$.
A convenient quantity which encodes the equilibration properties of
\emph{all }observables in $S$ is then given by the trace distance
between $\rho_{S}\left(t\right)$ and its equilibrium value $\overline{\rho_{S}}$,
i.e.\begin{equation}
D_{S}\left(t\right):=\left\Vert \rho_{S}\left(t\right)-\overline{\rho_{S}}\right\Vert _{1},\label{eq:trace-distance}\end{equation}
 where the norm-1 of a matrix $O$ is given by $\left\Vert O\right\Vert _{1}:=\frac{1}{2}\tr\sqrt{O^{\dagger}O}$.
The trace distance $D_{S}$ in Eq.~(\ref{eq:trace-distance}) was
first considered in \cite{linden09}. If the time average of $D_{S}\left(t\right)$
is small, the evolved state of the subsystem $\rho_{S}\left(t\right)$
stays close to its average $\overline{\rho_{S}}$ for most of the
time. This can be made quantitative by noting that $D_{S}$ is a positive
quantity and using Markov's inequality:\[
\mathrm{Prob}\left[D_{S}\left(t\right)\ge\epsilon\right]\le\frac{\overline{D_{S}}}{\epsilon},\]
 valid for any positive $\epsilon$. This implies that, for any observable
$O$ defined on $S$, the time evolved quantity $\langle O\left(t\right)\rangle$
is close to its mean $\overline{O}$ for most of the time. That is
if $\overline{D_{S}}$ is small, then one finds good equilibration
for all observables in $S$. This can be shown by noting that, if
the observable $O$ has its spectrum in $\left[o_{\mathrm{min}},o_{\mathrm{max}}\right]$
then \cite{linden09}:\begin{equation}
\left|\tr[\rho_{S}(t)O]-\tr[\overline{\rho_{S}}O]\right|\leq\left(o_{\mathrm{max}}-o_{\mathrm{min}}\right)D_{S}(t).\label{eq:trace_bound}\end{equation}
 The authors of \cite{linden09} also proved the following interesting
inequalities\begin{equation}
\overline{D_{S}}\le\frac{1}{2}\sqrt{\frac{d_{S}}{d^{\mathrm{eff}}\left(\overline{\rho}_{\overline{S}}\right)}}\le\frac{1}{2}\sqrt{\frac{d_{S}^{2}}{d^{\mathrm{eff}}\left(\overline{\rho}\right)}}.\label{eq:winter_inequality}\end{equation}
 Here $d_{S}$ is the dimension of subsystem $S$, equal to $2^{N_{S}}$
in our case, the effective dimension of a density matrix $d^{\mathrm{eff}}\left(\sigma\right)$
is defined as the inverse purity, i.e.~$d^{\mathrm{eff}}\left(\sigma\right)=1/\tr\left(\sigma^{2}\right)$.
Finally, $\overline{\rho}_{\overline{S}}$ is the equilibrium state
restricted to the complementary of $S$, i.e. its environment $\overline{S}$.
Especially the second inequality of Eq.~(\ref{eq:winter_inequality})
is very useful. It gives a bound on $\overline{D_{S}}$ by resorting
to a global quantity $d^{\mathrm{eff}}\left(\overline{\rho}\right)$.
As was shown in \cite{linden09,lcvpz1_2010,lcvpz2_2010}, for large
quenches $1/d^{\mathrm{eff}}\left(\overline{\rho}\right)$ is typically
exponentially small in the system size: $1/d^{\mathrm{eff}}\left(\overline{\rho}\right)\approx\exp\left(-\mathrm{const.}\times N\right)$.
There is, however, an interplay between system size and quench amplitude.
For moderate quenches, and/or when the system size is not too large,
the constant appearing in the exponential can be very small. This
can be seen using perturbation theory for small quenches which predicts
\cite{lcvpz1_2010,lcvpz2_2010,LCV07} $1/d^{\mathrm{eff}}\left(\overline{\rho}\right)\approx\exp\left(-\mathrm{const.}\times\delta h_{S}^{2}\times N_{S}\right)$.
Thus, when $N_{S}$ is not large compared to $(\delta h_{S})^{-2}$,
the second bound in Eq.~(\ref{eq:winter_inequality}) is not effective,
since $\overline{D_{S}}\le1$ by definition. Does this mean that equilibration
in this case will not take place in subsystem $S$? This is precisely
the question we are going to address here. Note that in our simulations
with $N_{S}=4$ the second bound in Eq.~(\ref{eq:winter_inequality})
is not effective as long as $1/d^{\mathrm{eff}}\left(\overline{\rho}\right)\ge0.015$,
whereas in most of our simulations we have $1/d^{\mathrm{eff}}\left(\overline{\rho}\right)>0.9$
(see below).

As a prototype quantity to monitor global equilibration of the entire
system, we study the Loschmidt echo (LE) \cite{jarzynski97,prosen98,kurchan00,pastawski01,quan06,rossini_base07,rossini07,PZ07,roux09}\[
\mathcal{L}\left(t\right)=\left|\langle\psi_{0}\vert\exp\left(-iHt\right)\vert\psi_{0}\rangle\right|^{2}.\]
 Here $H$ is the evolution Hamiltonian, and $|\psi_{0}\rangle$ is
the initial state. $\mathcal{L}\left(t\right)$ measures the probability
that the state $\vert\psi\left(t\right)\rangle$ evolves back to the
pre-quench initial state $\vert\psi_{0}\rangle$. In this sense, the
LE is the expectation value of a particular operator, i.e. the projector
onto the initial state $\vert\psi_{0}\rangle\langle\psi_{0}\vert$.
This operator has clearly support on the entire system, and therefore
the LE is the time dependent expectation value of a {}``global''
quantity. The LE has many interesting interpretations (see e.g.~\cite{silva08,lcvpz1_2010}
and reference therein). Most important for us, the time averaged LE
is precisely the inverse of the effective quantum dimension defined
above. In fact, writing the LE in the eigenbasis of $H$ one finds
\begin{equation}
\mathcal{L}\left(t\right)=\overline{\mathcal{L}}+2\sum_{n>m}p_{n}p_{m}\cos\left[t\left(E_{n}-E_{m}\right)\right],\label{eq:loschmidtexpansion}\end{equation}
 where $\overline{\mathcal{L}}=\sum_{n}p_{n}^{2}=\tr\left(\overline{\rho}|\psi_{0}\rangle\langle\psi_{0}|\right)=\tr\left[(\overline{\rho})^{2}\right]=1/d^{\mathrm{eff}}\left(\overline{\rho}\right)$.
Given its simple form in terms of the weights $p_{n}^{2}=\left|c_{n}\right|^{2}$,
the probability distribution of the LE is closely related to the distribution
of the $p_{n}$.

Once again, because of the bounds in Eqs.~(\ref{eq:winter_inequality})
and (\ref{eq:trace_bound}), a sufficiently small average of the LE
implies equilibration in all (sufficiently small) subsystems. Moreover,
using the bound $\overline{\mathcal{L}^{n}}\le n!\overline{\mathcal{L}}^{n}$
proved in \cite{lcvpz1_2010} with $n=2$ we obtain that the variance
satisfies $\overline{\Delta\mathcal{L}^{2}}\le\overline{\mathcal{L}}^{2}$,
and thus a small average implies small variance, with a signal to
noise ratio greater or equal to 1. This scenario can be considered
the standard road to equilibration. The converse need not be true,
i.e. one can have a small variance for the LE but a large mean. The
possibility of such alternative scenarios is what we are going to
explore in this paper.

\emph{Details of the numerical computation:} In the numerical computations
we take advantage of conservation of total magnetization to reduce
the Hilbert space dimension. To find the initial state $|\psi_{0}\rangle$,
we first diagonalize $H_{0}$ in the lowest magnetization sectors,
$S_{\mathrm{tot}}^{z}=0,1,2,3$, and keep the ground state. For the
quench parameters used below we verified that $|\psi_{0}\rangle$
always belongs to the zero magnetization sector. We then diagonalize
the evolution Hamiltonian $H$ using iterative Lanczos steps, calculating
the first 500 hundred lowest energy eigenstates. To obtain a good
approximation of the full sector we checked that the normalization
condition was satisfied up to $10^{-4}$, i.e.~$1-\sum_{n=1}^{500}p_{n}^{2}<10^{-4}$.
The time series statistics of observables are obtained using 400,000
random samples in a time window $\left[0,T_{\mathrm{max}}\right]$
where $T_{\mathrm{max}}$ is chosen more than two orders of magnitude
larger than the smallest energy gap in the system.

\begin{figure}[h]
\begin{centering}
\includegraphics[width=8cm]{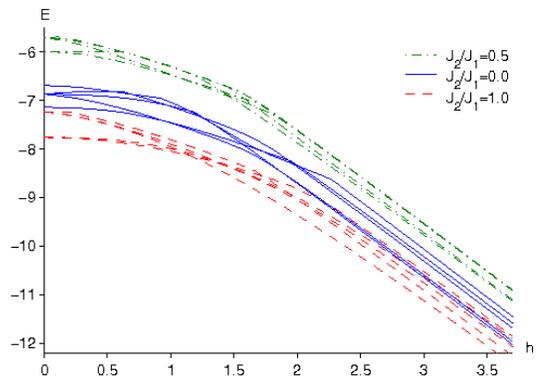}
\par\end{centering}

\caption{Lowest 5 energy levels of Hamiltonian Eq.~(\ref{eq:hamiltonian})
on a ring of $N=16$ sites for different coupling ratios $J_{2}/J_{1}$
as a function of the field $h_{S}$ applied on $N_{S}=4$ contiguous
sites. }

\label{fig:energydep}
\end{figure}

\section{Field-Energy-Dependence and Quenches in Different Regimes}

\label{sec:energydep} The Hamiltonian (\ref{eq:hamiltonian}) represents
a fully interacting quantum system. In order to gain some initial
insight into the physical properties of this system, we numerically
calculate the five lowest energy levels of the model as a function
of the applied local field $h_{S}$ on the four adjacent spins (representing
the subsystem S) in a chain of 16 spins for different ratios of the
nearest and next-nearest neighbor couplings (see Fig.~\ref{fig:energydep}).

\emph{Phase diagram at $h_{S}=0$:} The phase diagram for the model
at zero field and in the large $N$ limit is well known and has been
described for example in \cite{kwek}. At zero field and $J_{2}=0$
we have a Heisenberg spin chain with only nearest-neighbor coupling,
which is solvable using the Bethe Ansatz. For both positive $J_{1}$
and $J_{2}$ the system is frustrated. For small values of $J_{2}$
a gapless antiferromagnetic phase is present. At $J_{2}/J_{1}\approx0.241$
a gap opens up, and the system remains gapped for all finite $J_{2}/J_{1}$,
but the gap closes in the limit of $J_{2}/J_{1}\rightarrow\infty$,
where the model is approximately described by two weakly coupled chains.
At zero field and $J_{2}/J_{1}=1/2$ - the Majumdar-Ghosh point -
the ground state of the system factorizes and can be determined analytically.
Namely, the ground state manifold is two-fold degenerate spanned by
$|\phi_{1}\rangle,\,|\phi_{2}\rangle$ where $|\phi_{1}\rangle$ consists
of singlets between neighboring dimers, and $|\phi_{2}\rangle$ is
$|\phi_{1}\rangle$ translated by one site.

\emph{The model in a finite local field:} For small but finite local
fields - roughly $h_{S}<0.3$ - we encounter a regime, in which the
inter spin coupling dominates, and the local field acts as a perturbation.
The dependence of the eigenenergies on the local field here is relatively
flat. Degeneracies which occur for zero field are lifted.

For intermediate local fields - $0.5<h_{S}<2$ - the Heisenberg coupling
and the local magnetic field compete. In this regime we observe numerous
level crossings.

For large fields - $h_{S}>2.5$ - the energy levels are dominated
by the applied local magnetic field, and thus simply decrease linearly
with its amplitude. This is caused by the alignment of the affected
spins along the field direction, i. e. $m_{S}\rightarrow1/2$, thus
maximizing the contribution of the Zeeman energy term, $E_{Zeeman}=-h_{S}\cdot N_{S}/2$.
In this regime one can treat the Heisenberg interaction as a perturbation
on the field Hamiltonian for the spins with an applied field. In this
regime, the slope of all the energy levels in Fig.~\ref{fig:energydep}
approach what one would expect from a dominating Zeeman term $\frac{\partial E}{\partial h_{S}}\rightarrow N_{S}/2$.

Having these regimes of the model in mind, we consider the following
quench scenarios: small quenches within each of the regimes and large
quenches across different regimes.

\begin{figure}[t]
\begin{centering}
\includegraphics[width=9cm]{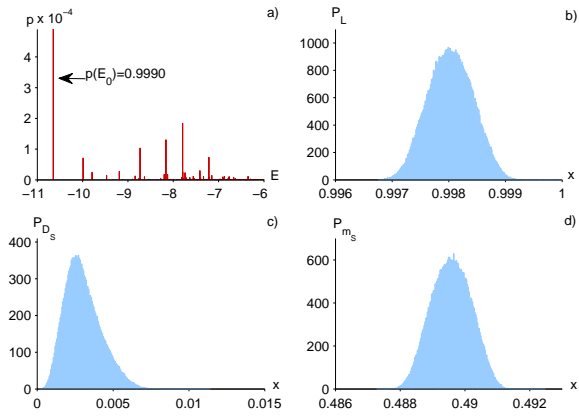}
\par\end{centering}

\caption{Equilibration statistics for $N=16$, $N_{S}=4$, $J_{2}/J_{1}=0$.
The quench is from from $h_{S}^{(i)}/J_{1}=3.5$ to $h_{S}^{(f)}/J_{1}=3$.
In panel a) the probabilities spectrum $p_{n}=p\left(E_{n}\right)$
is shown as a function of the energy. Panels b), c), d) show the full
statistics of $P_{\mathcal{L}},\, P_{D_{S}},\, P_{m_{S}}$ respectively.
Both the distribution of the LE and the magnetization are narrow Gaussian.
Moreover $P_{D_{S}}$ indicates a small mean $\overline{D_{S}}$.
All these facts indicate standard equilibration of Gaussian type.
Qualitatively similar results are observed for different values of
the ratio $J_{2}/J_{1}$.\label{fig:quench_gaussian} }

\end{figure}

The simplest case is a small quench within the regime of large dominating
local fields, where the energy levels in good approximation simply
vary linearly with the local field amplitude $h_{S}$. As an example
of this case, we consider the quench from an initial field $h_{S}^{(i)}=3.5$
to an evolution field $h_{S}^{(f)}=3$ for the three different couplings
$J_{2}/J_{1}=0,0.5,1$ (see Fig.~\ref{fig:quench_gaussian}). In
all of these cases the Loschmidt Echo as well as the local magnetization
show Gaussian distributions with very small variances, indicating
a good, straightforward. This is exactly the behavior expected for
small quenches in regular regimes \cite{lcvpz1_2010}. The distribution
of the quantity $D_{s}$ in these cases also resembles a Gaussian,
only experiencing a slight asymmetry due to the non-linearity of the
norm. So in regular regimes our example system shows good equilibration,
both locally and globally.

\begin{figure}[t]
\begin{centering}
\includegraphics[width=9cm]{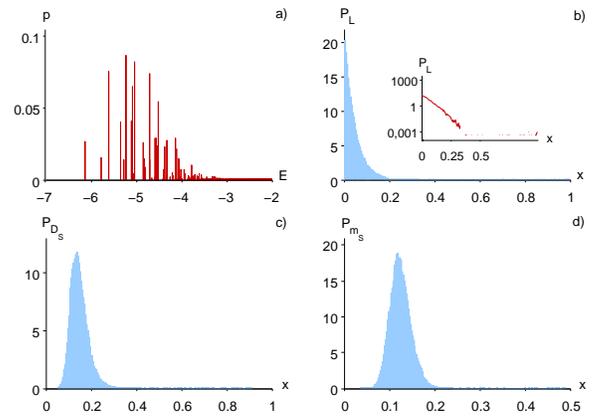}
\par\end{centering}

\caption{Equilibration statistics for $N=16$, $N_{S}=4$, $J_{2}/J_{1}=0$.
The quench is from from $h_{S}^{(i)}/J_{1}=3$ to $h_{S}^{(f)}/J_{1}=0$.
Panels as in figure \ref{fig:quench_gaussian}. As can be seen from
the log-linear plot in the inset of panel b), the distribution of
the LE is exponential with a very small mean. Qualitatively similar
results are observed for different values of the ratio $J_{2}/J_{1}$.}

\label{fig:quench_exponential}
\end{figure}

Another set of relatively simple quenches are those from large fields
to zero field. As the system in this case is strongly perturbed, one
would expect numerous excitations across a wide range of energies.
According to $\overline{\mathcal{L}}=1/d^{\mathrm{eff}}\left(\overline{\rho}\right)$
a large number of excitations causes a small Loschmidt Echo. More
specifically such quenches lead to an exponential distribution of
the Loschmidt Echo with an average very close to zero \cite{lcvpz1_2010}.
We numerically observe such a behavior when quenching from large to
zero local magnetic fields for all three couplings. Figure \ref{fig:quench_exponential}
shows a representative example of a quench from $h_{S}^{(i)}=3$ to
$h_{S}^{(f)}=0$ with only nearest-neighbor coupling. In this quench,
one also obtains single peaked and relatively narrow distributions
of the local magnetization and the quantity $D_{S}$. This indicates
measure concentration and exactly the kind of strong local equilibration
that is possible even for closed quantum systems which has been discussed
in \cite{linden09}. Results for different couplings are not shown
here, but very similar.

For small quenches in the regime of dominant Heisenberg couplings,
we observe different types of equilibration and a strong dependence
on the inter-spin coupling. This is discussed in the following section.

\section{Numerical Results for non-trivial quenches}

\label{sec:numres} Here we investigate small quenches of the form
$H_{0}(h_{S}^{(i)})\rightarrow H(h_{S}^{(f)})$ with $h_{S}^{(i)}=0.2$
and $h_{S}^{(f)}=0$ on systems of size $N=16$ and different coupling
ratio. In particular we present three simulation for quenches on a
subsystem $S$ of four spins ($N_{S}=4$) and coupling ratio $J_{2}/J_{1}=1,\,0,\,1/2$,~and
three analogous simulations for $N_{S}=3$. For all the simulations
the initial ground state is located in the $S_{\mathrm{tot}}^{z}=0$
sector of vanishing total magnetization.

\subsection{Quenches on four adjacent spins}

\begin{figure}[h]
\begin{centering}
\includegraphics[width=9cm]{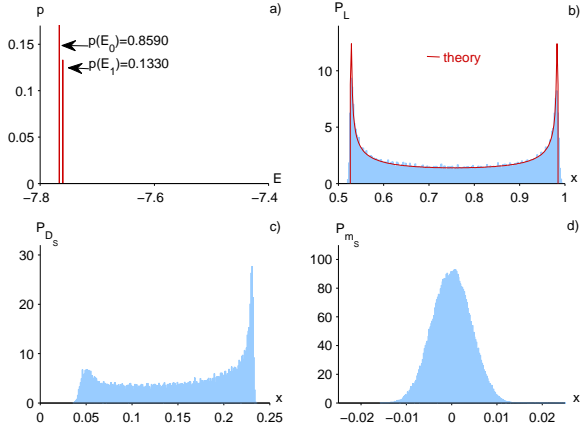}
\par\end{centering}

\caption{\emph{Quench 1}. Equilibration statistics for $N=16$, $N_{S}=4$,
$J_{2}/J_{1}=1$. The quench is from from $h_{S}^{(i)}/J_{1}=0.2$
to $h_{S}^{(f)}/J_{1}=0$. Panels as in figure \ref{fig:quench_gaussian}.}

\label{fig:small_lambda1}
\end{figure}

\emph{Quench 1:} The first example we consider is a quench on a system
of equal nearest and next-nearest neighbor couplings ($J_{1}=J_{2}$).
The system is quenched from $H_{0}(h_{S}^{(i)}=0.2)$ to $H(h_{S}^{(f)}=0)$.
The resulting equilibration statistics are shown in Fig.~\ref{fig:small_lambda1}.
As one would expect for a small quench, the energy probability distribution
$p(E_{n})$ in Fig.~\ref{fig:small_lambda1}a) is dominated by the
ground state ($p(E_{0})=0.86$). A more interesting feature is an
additional sizable contribution of the first excited state. Note that
the population of the first excited state is about two orders of magnitude
larger than the population of all the others ($p(E_{1})\ll p(E_{i}),i>1$).

The existence of these two dominating modes leads to a double-peaked
distribution function of the Loschmidt Echo $P_{\mathcal{L}}$, which
is clearly observed in Fig.~\ref{fig:small_lambda1}b). In this case,
one can effectively neglect the weights $p_{n}$ with $n\ge2$ in
Eq.~\ref{eq:loschmidtexpansion}, and treat the model as an effective
two-state system. Hence we obtain the following approximation for
the Loschmidt echo \begin{eqnarray}
\mathcal{L}(t) & \approx & \underbrace{p_{0}^{2}+p_{1}^{2}}_{\approx\overline{\mathcal{L}}}+2p_{0}p_{1}\cos[(E_{1}-E_{0})t].\label{eq:twomodl}\end{eqnarray}
 The corresponding probability distribution function can be calculated
analytically: \begin{equation}
P_{\mathcal{L}}(x)=\overline{\mathcal{L}}+\frac{1-\overline{\mathcal{L}}\left(x_{2}-x_{1}\right)}{\pi\sqrt{(x-x_{1})(x_{2}-x)}},\label{eq:double_peak}\end{equation}
 where $x_{1}=\overline{\mathcal{L}}-2p_{0}p_{1}$, $x_{2}=\overline{\mathcal{L}}+2p_{0}p_{1}$
denote the lower and upper edges of the distribution. Equation (\ref{eq:double_peak}),
with $p_{0}$ and $p_{1}$ obtained numerically, is shown as a red
line in Fig.~\ref{fig:small_lambda1}b) and shows excellent agreement
with the full calculation. This result shows that, after the quench,
the system oscillates between the two lowest states of $H$, $|0\rangle$
and $|1\rangle$, clearly indicating absence of equilibration. Such
a lack of equilibration can also be observed in the variance, $\Delta\mathcal{L}=\left(x_{2}-x_{1}\right)/\sqrt{8}$
which is of the same order of the support of the distribution. Especially
since the evolution Hamiltonian is translationally invariant, one
would also expect to see this non-equilibrium behavior locally. Accordingly,
the statistics of $D_{S}$ in Fig.~\ref{fig:small_lambda1}c) shows
lack of equilibration. Its distribution function also has a large
spread and two maxima, one of which shows a similar divergence as
the Loschmidt Echo. The asymmetry and the lack of a second peak are
not numerical artifacts but arise due to the high non-linearity of
the trace distance, namely its absolute value. A simplified example
of this behavior is given in the Appendix.

It is interesting to note that the distribution of the local magnetization
is Gaussian displaying good equilibration properties. The reason for
this becomes clear when looking at Eq.~(\ref{eq:observable}). The
largest (in modulus) contribution to that sum is given in principle
by the term with $n=1$ and $m=0$, but this contribution vanishes
as $\langle1|O|0\rangle=0$. Hence the sum in Eq.~(\ref{eq:observable})
contains a large number of terms of the same order of magnitude. In
this situation $\langle O\left(t\right)\rangle$ can be seen as a
sum of many independent variable converging to a Gaussian thanks to
a central limit theorem argument.

This situations makes clear that even when $\overline{D_{S}}$ is
large one can still find observable (those with very small or zero
weights in the low energy states) which show good equilibration properties
\footnote{For a quench using a field on only three adjacent spins, we observed
a similar $p_{n}$ distribution. Only here not the first, but the
second excited state gives a large contribution. This state has a
cross term with the ground state in the local magnetization of three
adjacent spins, and hence a double-peaked distribution function is
observed not only in the Loschmidt Echo, but also in the local magnetization.
(see Quench 4) %
}.

\begin{figure}[h]
\begin{centering}
\includegraphics[width=9cm]{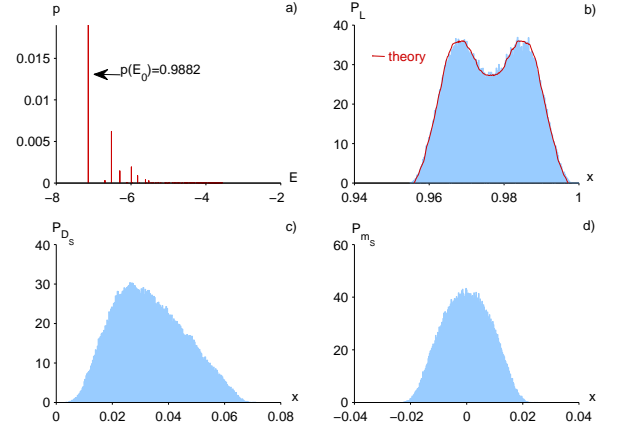}
\par\end{centering}

\caption{\emph{Quench 2}. Equilibration statistics for $N=16$, $N_{S}=4$,
$J_{2}/J_{1}=0$. The quench is from from $h_{S}^{(i)}/J_{1}=0.2$
to $h_{S}^{(f)}/J_{1}=0$. Panels as in figure \ref{fig:quench_gaussian}.
\label{fig:small_lambda0} }

\end{figure}

\emph{Quench 2:} The second quench we show is from a field of $h_{S}^{(i)}=0.2$
on four adjacent spins to zero field, but using only nearest-neighbor
coupling, i.e. $J_{2}=0$. The energy probability distribution of
this quench is shown in Fig.~\ref{fig:small_lambda0}a). In this
distribution the ground state is by far dominating ($p(E_{0})=0.99$),
the probability of the next highest populated state is two orders
of magnitudes smaller. Overall, there still is a concentration of
excitations in the low energies ($E<-5.5$).

As indicated by the distribution of $p(E_{n})$, the Loschmidt Echo
mean of this quench is much closer to one and its variance is about
an order of magnitude smaller compared with the case $J_{2}=J_{1}$
(Fig.~\ref{fig:small_lambda0}b)). Its distribution still shows two
maxima, but they are less pronounced and smoother.

In general, a better approximation to the LE distribution function
can be obtained by retaining the $N_{max}$ largest weights $W_{ij}=p_{i}p_{j}$
in equation (\ref{eq:loschmidtexpansion}). This procedure will be
performed for all of the following quenches. For the quench discussed
here, it is enough to keep $N_{max}=5$. The resulting approximate
distribution is shown as a red line in Fig \ref{fig:small_lambda0}b).

Both the LE and $D_{S}$ indicate a much better equilibration compared
to Quench 1. This is in contrast to the behavior of the local magnetization
in Fig.~\ref{fig:small_lambda1}d), which is again Gaussian, but
shows a larger rather than a smaller variance.

\begin{figure}[h]
\begin{centering}
\includegraphics[width=9cm]{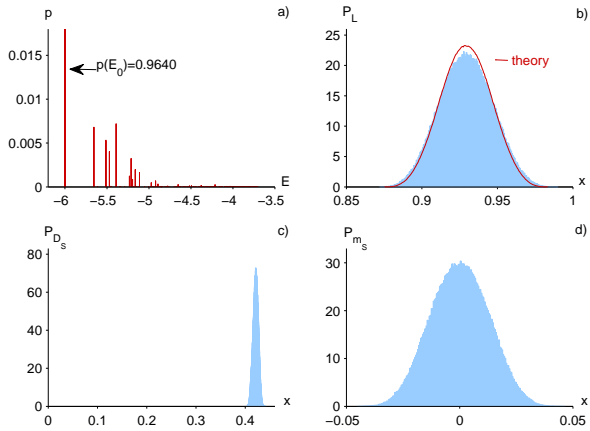}
\par\end{centering}

\caption{\emph{Quench 3}. Equilibration statistics for $N=16$, $N_{S}=4$,
$J_{2}/J_{1}=1/2$, Majumdar-Ghosh point. The quench is from from
$h_{S}^{(i)}/J_{1}=0.2$ to $h_{S}^{(f)}/J_{1}=0$. Panels as in figure
\ref{fig:quench_gaussian}. \label{fig:small_lambda_0.5} }

\end{figure}

\emph{Quench 3:} The last quench we want to discuss in detail is the
special case of $J_{2}=1/2$. The evolution system in zero field is
at the Majumdar-Ghosh point. The system has a two-fold degeneracy
at its minimum energy, consisting of two states, each of them a product
state of nearest-neighbor singlets. This degeneracy is lifted by the
pre-quench applied local field, with $h_{S}^{(i)}=0.2$ acting on
four adjacent spins. As a consequence the system cannot be treated
as non-degenerate. To take the degeneracies into account the corresponding
formulas have to be modified, e.g.~$\overline{\rho}$ includes off-diagonal
terms, where $E_{n}=E_{m}$ for $n\neq m$. Furthermore, because of
degeneracies, the proof for inequality (\ref{eq:winter_inequality})
given by \cite{linden09} does not hold in this case.

Inspecting $p(E_{n})$ in Fig.~\ref{fig:small_lambda_0.5}a), as
in the case of $J_{2}=0$ the population of the ground energy level
is by far dominating, $p(E_{0})=\tr\left[\Pi_{E_{0}}\rho_{0}\right]=0.96$
($\Pi_{E_{0}}$ is the projector onto the bi-dimensional ground state
manifold). We also observe many more excitations, none of which is
considerably larger than the others. The distribution is concentrated
at energies below $E=-5$, but even for $E>-5$ there are some weak
excitations. Quite a few of the lowest energy eigenstates, are not
populated at all, but protected by symmetry, e.g.~$p(E_{1})=p(E_{2})=0$.

Since the $p(E_{n})$ are relatively widely distributed, and besides
$p(E_{0})$ there is no other dominant contribution, we expect and
numerically obtain a Gaussian distributed LE, as shown in Fig.~\ref{fig:small_lambda_0.5}.
As in the previous quench we can approximate the LE truncating Eq.~\ref{eq:loschmidtexpansion}
to the first $N_{max}$ largest (in modulus) terms. As explained in
\cite{lcvpz2_2010}, in order to recover a Gaussian distribution we
have take $N_{max}$ large, here $N_{max}=20$ is needed. The corresponding
probability distribution is the line shown in Fig.~\ref{fig:small_lambda_0.5}b).

Compared to the quench at $J_{2}=0$ the LE mean is smaller and the
distribution shows a higher variance. So far, this indicates relatively
straightforward equilibration. Therefore the distribution of $D_{S}$
shown in Fig.~\ref{fig:small_lambda_0.5}c) is quite surprising.
It shows a slightly asymmetric Gaussian with the lowest variance of
all the quenches discussed, but with a relatively large mean $\overline{D_{S}}=0.419$.

A large $D_{S}$ does not mean bad equilibration for all local observables,
but it indicates that there is at least one local observable, which
equilibrates badly. As in the other quenches, the distribution function
of the local magnetization is a Gaussian centered around zero. Here,
in agreement with the LE, its variance is larger than in the case
$J_{2}=1$.

\subsection{Quenches on three adjacent spins}

\begin{figure}[h]
\begin{centering}
\includegraphics[width=9cm]{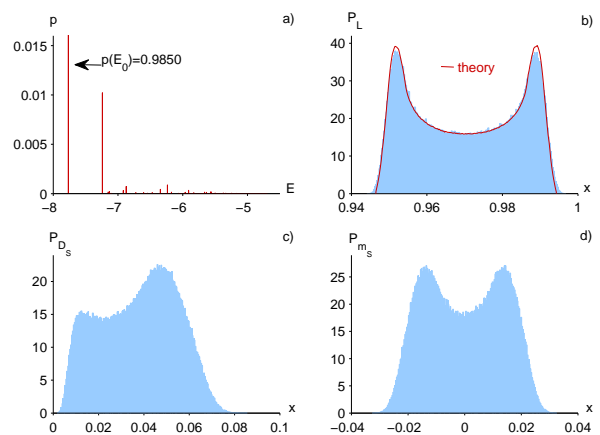}
\par\end{centering}

\caption{\emph{Quench} \emph{4}. Equilibration statistics for $N=16$, $N_{S}=3$,
$J_{2}/J_{1}=1$. The quench is from from $h_{S}^{(i)}/J_{1}=0.2$
to $h_{S}^{(f)}/J_{1}=0$. Panels as in figure \ref{fig:quench_gaussian}.
\label{fig:N3_small_lambda1} }

\end{figure}

\begin{figure}[h]
\begin{centering}
\includegraphics[width=9cm]{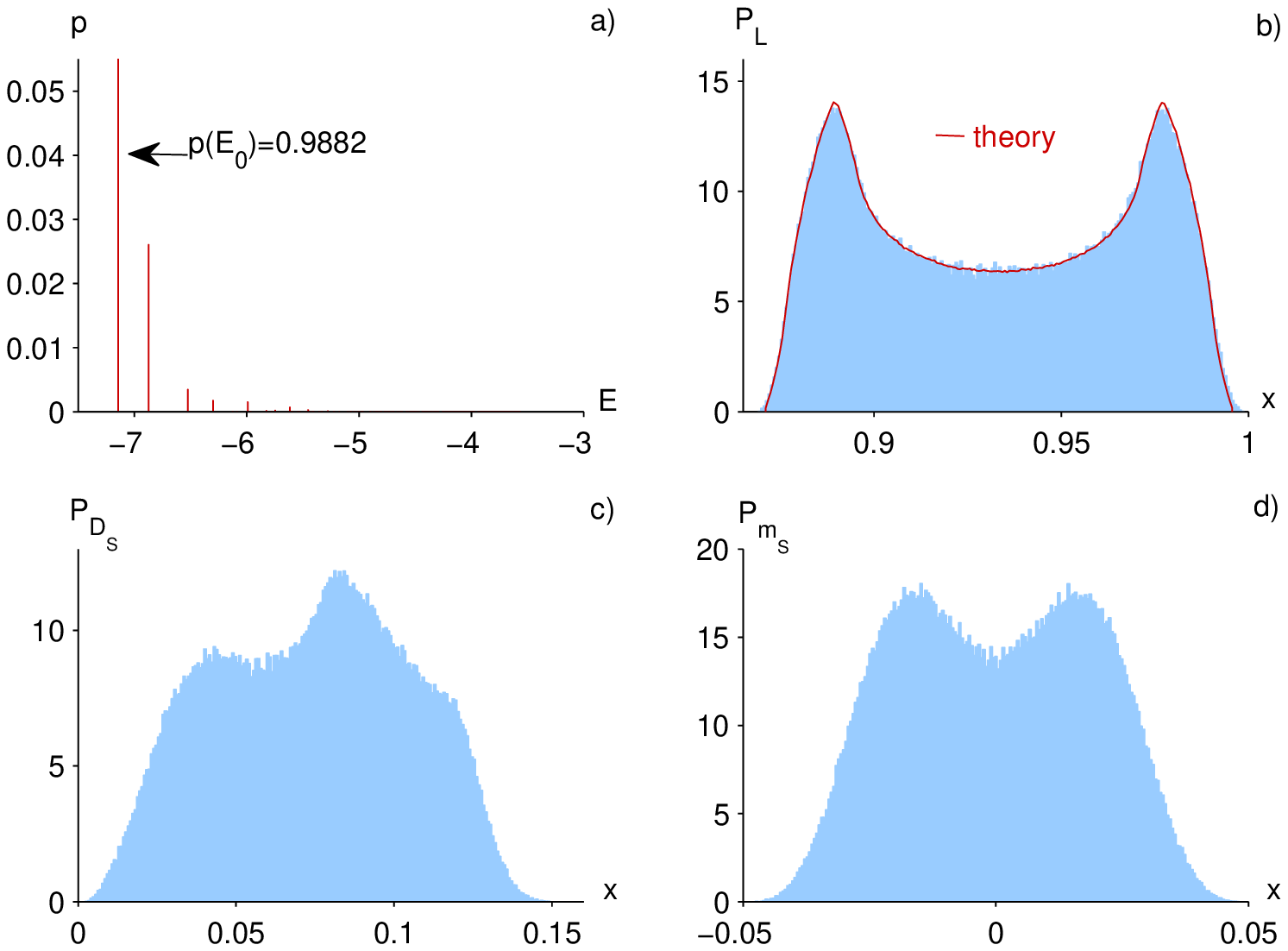}
\par\end{centering}

\caption{\emph{Quench }5. Equilibration statistics for $N=16$, $N_{S}=3$,
$J_{2}/J_{1}=0$. The quench is from from $h_{S}^{(i)}/J_{1}=0.2$
to $h_{S}^{(f)}/J_{1}=0$. Panels as in figure \ref{fig:quench_gaussian}.
\label{fig:N3_small_lambda0} }

\end{figure}

\begin{figure}[h]
\begin{centering}
\includegraphics[width=9cm]{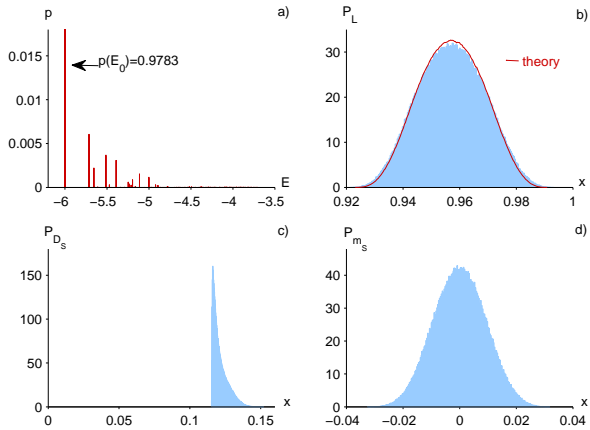}
\par\end{centering}

\caption{\emph{Quench 6. }Equilibration statistics for $N=16$, $N_{S}=3$,
$J_{2}/J_{1}=1/2$, Majumdar-Ghosh point. The quench is from from
$h_{S}^{(i)}/J_{1}=0.2$ to $h_{S}^{(f)}/J_{1}=0$. Panels as in figure
\ref{fig:quench_gaussian}. \label{fig:N3_small_lambda0.5} }

\end{figure}

We also show the equilibration statics of the same three quenches
just discussed, with the same quench amplitude but with a magnetic
field acting on three instead that on four adjacent spins, i.e.~$N=16$
and $N_{S}=3$ (Figs.~\ref{fig:N3_small_lambda1}-\ref{fig:N3_small_lambda0.5}).
These quenches do not need to be discussed in the same detail, but
indicate that some of the observed patterns are not restricted to
$N_{S}=4$ and provide a few interesting other features. Note also
that the corresponding variances of the considered observables $\mathcal{L}$,
$D_{S}$ and $m_{S}$ in the case of $N_{S}=3$ are much smaller than
for $N_{S}=4$.

\emph{Quench 4:} In the case of $J_{2}/J_{1}=1$ (Fig.~\ref{fig:N3_small_lambda1})
one obtains a similar dominance of two lowest-energy modes as for
the case $N_{S}=4$. Contributions of other states are only one order
of magnitude smaller than the two dominating modes. The LE distribution
accordingly is double peaked but smoother and with a wider spread
compared to the 4-site local quench that is about an order of magnitude
smaller. To obtain a good effective description using for the LE we
use here $N_{max}=5$. The contribution of these few other states
is even more visible in the distribution of $D_{S}$, which again
shows two maxima, but is much less spiked. Differently from the case
$N_{S}=4$, the local magnetization is double peaked following a distribution
function similar to the one of the LE. Correspondingly the variance
is roughly 4 times as large as for $N_{S}=4$, indicating worse equilibration.
This can simply be explained by the fact that the two dominant modes,
namely the ground state and the second excited state in this case
have a finite cross term in the local magnetization.

\emph{Quench 5:} The quench at $J_{2}/J_{1}=0$ shown in Fig.~\ref{fig:N3_small_lambda0}
displays similar features as in the case $J_{2}/J_{1}=1$, with all
the distributions that tend to be broader.

\emph{Quench 6:} The last quench shown in Fig.~\ref{fig:N3_small_lambda0.5}
is the case $J_{2}/J_{1}=1/2$, where the evolution Hamiltonian is
at the Majumdar-Ghosh point. As in the case for $N_{S}=4$ one observes
a relatively large number of non-vanishing modes and a Gaussian distribution
of both the LE and the local magnetization. The approximation shown
in Fig.~\ref{fig:N3_small_lambda0.5}b) is obtained using $N_{max}=10$.
Although the distribution of $D_{S}$ is highly peaked with a small
variance its average is relatively large, indicating that there can
be local observables experiencing poor equilibration.

\section{Conclusions}

In this paper we have numerically studied different local quantum
quenches on frustrated spin chains. More precisely we considered both
global and local equilibration indicators, the Loschmidt echo and
the subsystem magnetization. To gain a better understanding of the
equilibration properties of the subsystem we also studied the distance
of the reduced density matrix from its time average: $D_{S}\left(t\right)=\left\Vert \rho_{S}\left(t\right)-\overline{\rho_{S}}\right\Vert $.
Our numerical simulations show that a variety of possible scenarios
take place depending on the interplay between quench amplitudes and
coupling constants. In particular, for large and intermediate quenches
the long time probability distribution of the Loschmidt echo resemble
an exponential and a Gaussian respectively, as it was first observed
in \cite{lcvpz1_2010}. When the Loschmidt echo mean is very small
the subsystem does equilibrate even when the bounds in \cite{linden09}
are too loose to be applicable. On the contrary, when the Loschmidt
echo mean is large and thus no general conclusion can be drawn a priori,
we observe all possible situations. Namely we observe both cases where
the subsystem equilibrates (\emph{quenches 2, 4})\emph{ }and where
it does not (\emph{quenches 1, 3, 5, 6}). Moreover even when the subsystem
does not equilibrate as a whole, we still encounter instances where
one observable (the magnetization) shows good equilibration properties
(\emph{quenches 1, 3}).

In conclusion, we would like to note that the identification and
characterization of the time-scales associated with the different
equilibration phenomena we observed lies outside the scope of the
long-time statistics techniques adopted in this paper. Addressing
this problem in its generality is perhaps the most exciting next
challenge in the quest for understanding equilibration dynamics of
unitarily evolving quantum systems.

P.Z.~acknowledges support from NSF grants PHY-803304, DMR-0804914
and L.C.V.~acknowledges support from European project COQUIT under
FET-Open grant number 2333747.

\bibliographystyle{apsrev}
\bibliography{loc_que}

\appendix

\section*{Appendix}

Let us assume that the system is divided into a subsystem $S$ and
an environment $E$. The two corresponding bases are given by $\{\left|a\right>,\left|b\right>,\cdots\}$
and $\{\left|\alpha\right>,\left|\beta\right>,\cdots\}$. To model
quench 1 and to simplify the calculation we further assume that only
two eigenstates of the closed system $S\otimes E$ contribute to the
full density matrix. (A reasonable simplification of the $p_{n}$
distribution in Fig. \ref{fig:small_lambda1}a).) To model the strongly
coupled system and to obtain a non-unitary evolution of the subsystem,
we have to introduce entanglement. This can be done by considering
the two states \begin{eqnarray}
\left|1\right> & \equiv & \left(|a,\alpha\rangle+|b,\beta\rangle\right)/\sqrt{2},\\
\left|2\right> & \equiv & \left(|a,\alpha\rangle-|b,\beta\rangle\right)/\sqrt{2}.\label{eq:onetwo}\end{eqnarray}
 Using the initial state $\left|\Psi_{0}\right>\equiv c_{1}\left|1\right>+c_{2}\left|2\right>$
we compute $D_{S}(t)$ and its distribution. Given the energy difference
$\omega=E_{2}-E_{1}$ of the two eigenstates we obtain \begin{eqnarray}
\rho-\overline{\rho} & = & c_{1}c_{2}^{*}\left|1\right>\left<2\right|e^{-i\omega t}+h.c.\label{eq:rhor}\end{eqnarray}
 Tracing out the degrees of freedom of the bath this simplifies to
\begin{eqnarray}
\rho_{S}-\overline{\rho_{S}} & = & \Tr_{B}\left(\rho-\overline{\rho}\right)\\
 & = & c_{1}c_{2}^{*}\left(\left|a\right>\left<a\right|-\left|b\right>\left<b\right|\right)e^{i\omega t}+h.c..\label{eq:rhosr}\end{eqnarray}
 Identifying $c_{1}c_{2}^{*}=\sqrt{p_{1}p_{2}}e^{i\varphi}$ and using
a matrix representation we get \begin{eqnarray}
\rho_{S}-\overline{\rho_{S}} & = & \sqrt{p_{1}p_{2}}\cos\left(\omega t+\varphi\right)\left(\begin{array}{rr}
1 & 0\\
0 & -1\end{array}\right)\label{eq:rhosr2}\end{eqnarray}
 We finally obtain \begin{eqnarray}
D_{S}(t) & = & \sqrt{p_{1}p_{2}}\left|\cos\left(\omega t+\varphi\right)\right|\label{eq:dsexm}\end{eqnarray}
 Its distribution function can be calculated analytically, giving
a simple description of a divergence similar to what we observe numerically,
\begin{eqnarray}
P_{D_{S}}(x) & = & \lim\limits _{T\rightarrow\infty}\frac{1}{T}\int_{0}^{T}\delta\left(D_{S}\left(t\right)-x\right)dt\\
 & = & \frac{\omega}{\pi}\int_{0}^{\frac{\pi}{\omega}}\delta\left(\sqrt{p_{1}p_{2}}\left|\cos\left(\omega t\right)\right|-x\right)dt\\
 & = & \frac{2/\pi}{\sqrt{p_{1}p_{2}-x^{2}}}.\label{eq:pdsexm}\end{eqnarray}
 Note that in this case $D_{S}$ takes values in $D_{S}\in\left[0,\sqrt{p_{1}p_{2}}\right]$,
so such a $P\left(d_{S}\right)$ has only one square root singularity
at the upper edge. If we simply take the two largest amplitude contributions
of Fig.~\ref{fig:small_lambda1}a), we can estimate the peak in Fig.~\ref{fig:small_lambda1}c)
to be around $\sqrt{0.86\cdot0.13}=0.33$. Given the great simplification
this estimate is quite useful.
\end{document}